\documentclass[conference]{IEEEtran}
\usepackage{cite}
\usepackage{graphicx}
\usepackage{psfrag}
\usepackage{subfigure}
\usepackage{flushend}
\usepackage{color}

\usepackage{amsmath}
\usepackage{array}
\usepackage{graphics}
\usepackage{epsfig}
\usepackage{amsbsy}
\usepackage{amssymb}
\usepackage{amsthm}

\usepackage{framed}
\usepackage{algorithm}
\newtheorem{proposition}{Proposition} 
\IEEEoverridecommandlockouts

\begin{document}

\title{\LARGE Interference-Assisted Wireless Energy Harvesting in\\ Cognitive Relay Network with Multiple Primary Transceivers\vspace*{-2mm}}
\author{Sanket S.~Kalamkar\IEEEauthorrefmark{1} and~Adrish~Banerjee\vspace*{-2mm}
\thanks{The authors are with the Department of Electrical Engineering, Indian Institute of Technology, Kanpur, India (e-mail: kalamkar@iitk.ac.in, adrish@iitk.ac.in).}
\thanks{\IEEEauthorrefmark{1}The author is supported by the TCS research scholarship.}

}

\maketitle

\begin{abstract}
We consider a spectrum sharing scenario, where a secondary network coexists with a primary network of multiple transceivers. The secondary network consists of an energy-constrained decode-and-forward secondary relay which assists the communication between a secondary transmitter and a destination in the presence of the interference from multiple primary transmitters. The secondary relay harvests energy from the received radio-frequency signals, which include the information signal from the secondary transmitter and the primary interference. The harvested energy is then used to decode the secondary information and forward it to the secondary destination. At the relay, we adopt a time switching policy due to its simplicity that switches between the energy harvesting and information decoding over time. Specifically, we derive a closed-form expression for the secondary outage probability under the primary outage constraint and the peak power constraint at both secondary transmitter and relay. In addition, we investigate the effect of the number of primary transceivers on the optimal energy harvesting duration that minimizes the secondary outage probability. By utilizing the primary interference as a useful energy source in the energy harvesting phase, the secondary network achieves a better outage performance. 
\end{abstract}

\section{Introduction}
Energy harvesting (EH) cognitive radio~\cite{sultan,lee1,jeya,usman,shaf} is a promising solution to the problem of the inefficient spectrum usage while achieving green communications. In particular, the cognitive radio can improve the spectral efficiency by facilitating the unlicensed/secondary users (SUs) to share the spectrum with the licensed/primary users (PUs), provided that the interference to PUs stays below a specified threshold. On the other hand, energy harvesting provides the cognitive radio a greener alternative to harness energy for its operation, which also helps enhance its lifetime under the energy constraints.

Besides harvesting energy from natural sources like solar and wind, nowadays, the radio environment can feed the energy in the form of radio-frequency (RF) signals~\cite{lu}. Noticing that RF signals can carry both information and energy together, \cite{varshney,grover,rui2} have advocated the use of RF signals to harvest energy along with the information transmission. However, it is difficult for a receiver, in practice, to simultaneously decode the information and harvest energy from the received RF signals. Thus, two practical policies are proposed to harvest energy and decode information separately~\cite{rui2,rui3, nasir}. One is the time switching policy, where the time is switched between the energy harvesting and information decoding; while the second policy is based on the power splitting, where a part of the received power is used to harvest energy and the rest for information decoding. 

Such wireless energy harvesting while receiving the information has an important application in cooperative relaying, where an intermediate node helps forwarding the information from the source to the destination to improve the coverage and reliability of the communication~\cite{nasir,aissa,krik,yener1,ishibashi1,poor,gan,micha,nasir2,chen1}. However, the relay may have a battery with limited capacity, replacing or recharging which frequently may be inconvenient. In this case, wireless energy harvesting helps the relay to stay active in the network and facilitate the information cooperation. Similarly, in cognitive radio, using energy harvesting for energy-limited relays, SUs can achieve significant performance gains~\cite{van,mousa}. In \cite{mousa}, under spectrum sharing with a PU, an EH relay which forwards the secondary data is considered, while a tradeoff between primary interference constraint and energy constraint due to EH nature of relays is investigated in~\cite{sanket}.

In spectrum sharing, both PU and SU transmit together, which limits the transmit powers of secondary source and relay to keep the interference to PU below a threshold. However, PU, being a legacy user, has no such restriction on its transmit power. Due to this, SU may experience heavy interference from PU, which deteriorates the quality-of-service (QoS) of SU. Nevertheless, since the interference is a RF signal, it can be leveraged as a potential source of energy~\cite{rui3,aissa,he}. For example, under time switching policy, in the energy harvesting phase of a slot, the interference can be utilized as a useful energy source. This could subdue the harmful effect of the interference at the energy-constrained relay by supplying additional energy, which can be used to transmit with a higher power (provided it satisfies PU's interference threshold), to achieve better QoS.

The contributions and key results of this paper are as follows:
\begin{itemize}
\item With interference leveraged as an energy source, under spectrum sharing with multiple primary transceivers, we consider SU's communication via a decode-and-forward relay that harvests energy from the received RF signals, i.e., the information signal from the secondary source and the primary interference, using the time switching policy.  
\item For the proposed model, we derive a closed-form expression for SU's outage probability provided PU's outage probability remains below a threshold and investigate the effective use of the interference from multiple primary transmitters as an energy source.
\item We show that, such interference-assisted EH not only improves SU's outage performance due to the extra acquired energy, but also reduces the optimal energy harvesting time that minimizes the secondary outage probability.
\item  Finally, we study the impact of the number of primary transceivers on SU's outage performance. We observe that, though the optimal energy harvesting time reduces with the increase in the number of primary transceivers, the minimum secondary outage probability increases simultaneously when the transmit powers of the secondary transmitter and relay restricted by the primary outage constraint satisfy the peak power constraint. Interestingly, the trend reverses once the peak power constraint limits the transmit powers of the secondary transmitter and the relay.\vspace*{-1mm}
\end{itemize}
\begin{figure}
\centering
\includegraphics[scale=0.21]{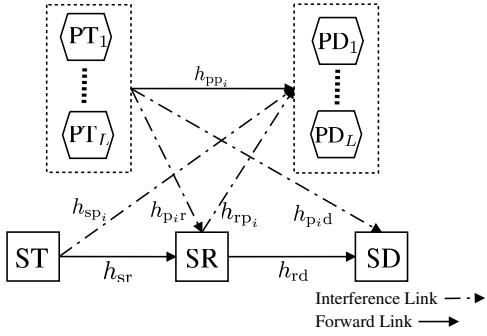}\vspace*{-2mm}
\caption{Secondary communication via an EH relay in spectrum sharing.}
\label{fig:syst}\vspace*{-4mm}
\end{figure}

\section{System and Channel Models}
As shown in Fig.~\ref{fig:syst}, consider a primary network consisting of $L$ pairs of primary transmitters (PTs) and primary destinations (PDs), where each PU pair communicates over a channel of bandwidth $B$~$\mathrm{Hz}$. The secondary network consists of a secondary transmitter (ST) which communicates with a secondary destination (SD) through an energy harvesting decode-and-forward secondary relay (SR). All nodes have a single antenna. The secondary network shares the spectrum of bandwidth $BL$~$\mathrm{Hz}$ with PUs, provided that the quality-of-service (QoS) of each primary link is maintained above a given threshold.

Let $h_{{\mathrm{pp}}_{i}}$, $h_{{\mathrm{sr}}}$, $h_{{\mathrm{rd}}}$, $h_{{\mathrm{sp}}_{i}}$, $h_{{\mathrm{rp}}_{i}}$, $h_{{\mathrm{p}}_i\mathrm{r}}$, and $h_{{\mathrm{p}}_i\mathrm{d}}$ denote the channel coefficients of $i$th primary link PT$_{i}$-PD$_{i}$ ($i = 1, 2, \dotsc, L$), ST-SR, SR-SD, ST-PD$_i$, SR-PD$_{i}$, PT$_i$-SR, and PT$_i$-SD, respectively. All channels are independent of each other and experience quasi-static Rayleigh fading, i.e., the channels remain constant for one slot of secondary communication and change independently from one slot to another. The instantaneous channel power gains are exponentially distributed random variables (RVs). Let us denote the mean channel power gain of $|h_{\mathrm{k}}|^{2}$ by $\lambda_{\mathrm{k}}$, where $\mathrm{k} \in \lbrace{{\mathrm{pp}}_{i}, \mathrm{sr}}, {\mathrm{rd}}, {\mathrm{sp}}_{i}, {\mathrm{rp}}_{i}, {\mathrm{p}}_i\mathrm{r},  {\mathrm{p}}_i\mathrm{d}\rbrace$. For simplicity, we consider PT-PR links are identically distributed, i.e., $\lambda_{{\mathrm{pp}}_{i}}= \lambda_{{\mathrm{pp}}}$; interference channels from PTs to a node and vice-versa are also identically distributed, i.e., $\lambda_{{\mathrm{p}}_i\mathrm{r}} = \lambda_{{\mathrm{p}}\mathrm{r}}$, $\lambda_{{\mathrm{p}}_i\mathrm{d}} = \lambda_{{\mathrm{p}}\mathrm{d}}$, $\lambda_{{\mathrm{s}\mathrm{p}}_i} = \lambda_{{\mathrm{s}}\mathrm{p}}$, and  $\lambda_{{\mathrm{r}\mathrm{p}}_i} = \lambda_{{\mathrm{r}}\mathrm{p}}$. We assume the knowledge of mean channel power gains for PT$_i$-PD$_i$, ST-PD$_i$, and SR-PD$_i$ links, while SR and SD have the knowledge of instantaneous channels gains for the respective receiving links, i.e., for ST-SR and PT$_i$-SR links at SR and for SR-SD and PT$_i$-SD links at SD, as in~\cite{zou,peter:2013}.

Assuming no direct link between transmitter and destination due to high attenuation~\cite{aissa, nasir,krik,yener1}, the secondary communication happens over two-hops. In the first hop, ST transmits data to SR, while in the second hop, SR forwards the received data to SD after decoding. The SR is an EH node, that is capable of harvesting energy from the received radio-frequency (RF) signals. Energy harvesting is considered to be the only power source for SR. The SR may use some part of the received information signal to gather the energy required to forward the information to SD. In addition, as in spectrum sharing, the primary and secondary network transmit simultaneously, SR experiences the interference from $L$ PTs, which is also a RF signal. Thus, SR can also harvest additional energy from the primary interference in the energy harvesting phase, converting it into a useful energy source. The ST and PTs are the conventional nodes with constant power supply (e.g. battery).

\section{Maximum Allowed Secondary Transmit Powers}
In the spectrum sharing scenario, the interference constraints at PDs govern the maximum transmit powers of ST and SR. We model the interference constraint at a PD as its outage probability, i.e., ST and SR should limit their transmit powers so that the outage probability of each primary link remains below a given threshold. Let us denote the maximum allowed transmit powers of ST and SR due to the primary outage constraint as $P_{\mathrm{ST}}$ and $P_{\mathrm{SR}}$, respectively. Then, in the first hop of the secondary communication when ST transmits to SR, given the constant transmit power of PT ($P_{\mathrm{PT}}$), the outage probability for $i$th primary link can be written as follows:\vspace*{-1mm}
\begin{equation}
\mathrm{P}^{i}_{\mathrm{p, out, ST}} = \mathrm{Pr}\left(B\log_{2}\left(1+ \gamma_{\mathrm{PD}_i}\right)\leq\mathcal{R}_\mathrm{p}\right) \leq \Theta_{\mathrm{p}},
\label{eq:P_out_st}\vspace*{-1mm}
\end{equation}
where $\gamma_{\mathrm{PD}_i} = \frac{P_{\mathrm{PT}}|h_{\mathrm{pp}_i}|^{2}}{P_{\mathrm{ST}}|h_{\mathrm{sp}_i}|^{2}}$ is the signal-to-interference ratio (SIR)\footnote{Since our focus is interference-limited spectrum sharing environment where the interference power is dominant than the noise power, the latter can be neglected~\cite{duong2}.} at PD$_{i}$, $\mathcal{R}_\mathrm{p}$ is the desired primary rate for each primary link, and $\Theta_{\mathrm{p}}$ is the primary outage threshold for each PU. Ensuring that the outage probability of the primary link having the worst SIR stays below $\Theta_{\mathrm{p}}$, we can write the primary outage constraint with interference from ST as\vspace*{-1mm}
\begin{equation}
\mathrm{P}_{\mathrm{p, out, ST}} = \mathrm{Pr}\left(\max_{i = 1, 2, \dotsc, L} \mathrm{P}^{i}_{\mathrm{p, out, ST}}\right) \leq \Theta_{\mathrm{p}}.
\label{eq:Ppout}\vspace*{-1mm}
\end{equation}
Then, from \eqref{eq:P_out_st}, and using the independence between $|h_{\mathrm{pp}_i}|^{2}$ and $|h_{\mathrm{sp}_i}|^{2}$, we can write \eqref{eq:Ppout} as\vspace*{-1mm}
\begin{align}
\mathrm{P}_{\mathrm{p, out, ST}}  = 1 - \prod_{i = 1}^{L} \left(1-\mathrm{Pr}\left( \frac{P_{\mathrm{PT}}|h_{\mathrm{pp}_i}|^{2}}{P_{\mathrm{ST}}|h_{\mathrm{sp}_i}|^{2}} \leq \zeta_{\mathrm{p}}\right)\right),
\label{eq:Ppout2}
\end{align}
where $\zeta_{\mathrm{p}} = 2^{\mathcal{R}_\mathrm{p}/B} - 1$.\vspace*{-1mm}
\begin{proposition}
The maximum allowed transmit power for ST under the primary outage constraint is\vspace*{-1mm}
\begin{equation}
P_{\mathrm{ST}} = \frac{P_{\mathrm{PT}}\lambda_{\mathrm{pp}}}{\zeta_{\mathrm{p}} \lambda_{\mathrm{sp}}}\left(\left(\frac{1}{1-\Theta_{\mathrm{p}}}\right)^{\frac{1}{L}} - 1\right)^{+},
\label{eq:PST}\vspace*{-1mm}
\end{equation}
where $(x)^{+} = \max(x, 0)$.
\end{proposition}\vspace*{-2mm}
\begin{proof}
The proof is given in Appendix \ref{sec:PST}.
\end{proof}

Similarly, in the second hop of the secondary communication when SR transmits to SD, following the same procedure to derive $P_{\mathrm{ST}}$, the maximum transmit power for SR is given as\vspace*{-1mm}
\begin{equation}
P_{\mathrm{SR}} = \frac{P_{\mathrm{PT}}\lambda_{\mathrm{pp}}}{\zeta_{\mathrm{p}} \lambda_{\mathrm{rp}}}\left(\left(\frac{1}{1-\Theta_{\mathrm{p}}}\right)^{\frac{1}{L}} - 1\right)^{+}.
\label{eq:Ppout7}\vspace*{-1mm}
\end{equation}
Besides the primary outage constraint, at both ST and SR, we also impose the peak power constraint $P_{\mathrm{t}}$. Then, the maximum transmit powers for ST and SR respectively become\vspace*{-1mm}
\begin{equation}
P_{\mathrm{Sm}} = \min\left(P_{\mathrm{ST}}, P_{\mathrm{t}}\right),
\label{eq:ST}
\end{equation} 
\begin{equation}
P_{\mathrm{R}} = \min\left(P_{\mathrm{SR}}, P_{\mathrm{t}}\right).
\label{eq:SR}\vspace*{-3mm}
\end{equation}

\begin{figure}
\centering
\includegraphics[scale=0.26]{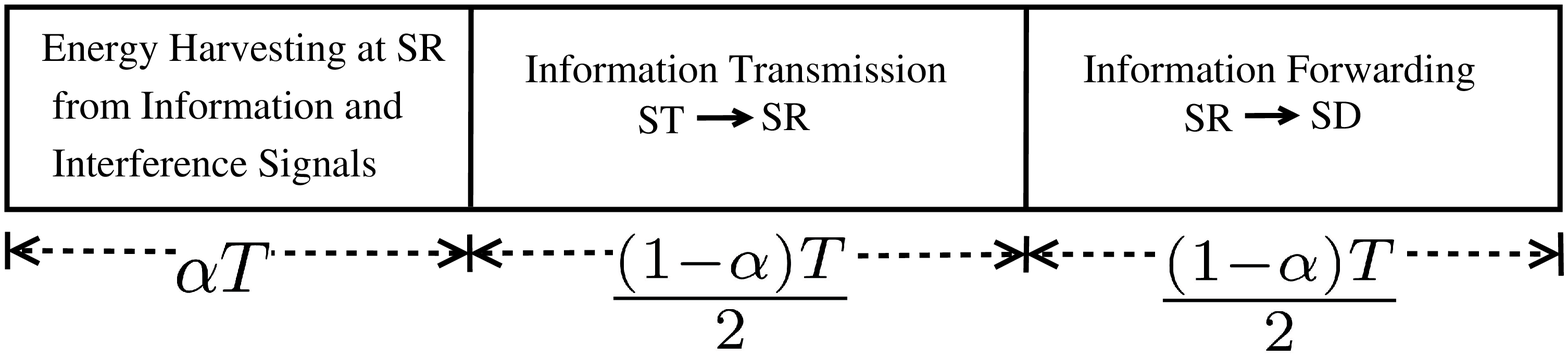}\vspace*{-2mm}
\caption{Time switching protocol for the interference-assisted energy harvesting and information processing at SR.}\vspace*{-4mm}
\label{fig:protocol}
\end{figure}

\section{Relaying Protocol at Secondary Relay}
\label{sec:relay_prot}

In this paper, at SR, we adopt a time switching protocol due to its simplicity to harvest energy from received RF signals as shown in Fig.~\ref{fig:protocol}. In this protocol, at the start of a slot, for $\alpha T$ duration ($0 < \alpha < 1$), SR harvests energy from ST's signal and interference from $L$ PTs, where $T$ is the duration of one slot of the secondary communication. The remaining time slot of duration $(1-\alpha)T$ is divided into two sub-slots, each of duration $\frac{(1-\alpha)T}{2}$. In the first sub-slot, ST transmits information to SR; while SR forwards the information to SD in the next sub-slot. Thus, when ST transmits with $P_{\mathrm{Sm}}$ and each PT transmits with $P_{\mathrm{PT}}$, the energy harvested by SR in $\alpha T$ duration is given as\vspace*{-1mm}
\begin{equation}
E_{\mathrm{SR,H}} = (\alpha T)\delta \left(P_{\mathrm{Sm}}|h_{\mathrm{sr}}|^{2} + \sum_{i = 1}^{L} P_{\mathrm{PT}}|h_{{\mathrm{p}}_i\mathrm{r}}|^2  \right),
\label{eq:harv_energy}\vspace*{-1mm}
\end{equation}
where $\delta$, with $0 \leq \delta \leq 1$, is the energy conversion efficiency factor, whose value depends on the receiver architecture. The SR uses the harvested energy to forward the information to SD. Then, given the amount of harvested energy, the transmit power of SR in the absence of peak power constraint and primary outage constraint can be given by\footnote{Usually, the energy consumption by the circuitry of SR in the information processing is negligible compared to that in the transmission~\cite{nasir,nasir2}. Thus, we assume that SR uses all the harvested energy for the transmission.}\vspace*{-1mm}
\begin{equation}
\!P_{\mathrm{SR, H}} = \frac{2 E_{\mathrm{SR,H}}}{(1-\alpha)T} = \frac{2\delta \alpha}{1-\alpha}\!\left(\!\!P_{\mathrm{Sm}}|h_{\mathrm{sr}}|^{2} + \! \sum_{i = 1}^{L}P_{\mathrm{PT}}|h_{{\mathrm{p}}_i\mathrm{r}}|^2 \!\right).
\label{eq:harv_power}\vspace*{-1mm}
\end{equation}

\noindent Now, by incorporating the primary outage constraint and the peak power constraint, the maximum transmit power for the energy harvesting SR can be given as follows:\vspace*{-2mm}
\begin{equation}
P_{\mathrm{Rm}} = \min\left(P_{\mathrm{SR, H}}, P_{\mathrm{R}}\right),
\label{eq:SR_fin}\vspace*{-1mm}
\end{equation}
where $P_{\mathrm{R}}$ is given by \eqref{eq:SR}. Hereafter, without loss of generality, we assume that the duration of a time-slot is $T = 1$.\vspace*{-1mm}

\section{Secondary Outage Analysis}
The secondary communication between ST and SD via SR experiences an outage if the rate on one of the ST-SR and SR-SD links falls below the desired rate $\mathcal{R}_{\mathrm{s}}$. Then, we can write the secondary outage probability $P_{\mathrm{s,out}}$ as follows:
\begin{equation}
P_{\mathrm{s,out}} = \mathrm{Pr}\left(\min \left(R_{\mathrm{sr}}, R_{\mathrm{rd}}\right) < \mathcal{R}_{\mathrm{s}}\right),
\label{eq:Psout}\vspace*{-1mm}
\end{equation}
where $R_{\mathrm{sr}}$ and $R_{\mathrm{rd}}$ are the rates on ST-SR and SR-SD links, respectively, and can be given as
\begin{align}
R_{\mathrm{sr}} &= \frac{1-\alpha}{2}BL\log_{2}\left(1 + \gamma_{\mathrm{SR}}\right),\nonumber\\
R_{\mathrm{rd}} &= \frac{1-\alpha}{2}BL\log_{2}\left(1 + \gamma_{\mathrm{SD}}\right).
\end{align}
Here, $\gamma_{\mathrm{SR}}$ and $\gamma_{\mathrm{SD}}$ are SIRs at SR and SD, respectively, and are given as\vspace*{-1mm}
\begin{equation}
\gamma_{\mathrm{SR}} = \frac{P_{\mathrm{Sm}} |h_{\mathrm{sr}}|^{2}}{\displaystyle \sum_{i = 1}^{L}P_{\mathrm{PT}}|h_{\mathrm{p}_i \mathrm{r}}|^2},
\label{eq:SIR_SR}\vspace*{-1mm}
\end{equation}
\begin{equation}
\gamma_{\mathrm{SD}} = \frac{P_{\mathrm{Rm}} |h_{\mathrm{rd}}|^{2}}{\displaystyle \sum_{i = 1}^{L}P_{\mathrm{PT}}|h_{\mathrm{p}_i \mathrm{d}}|^2}.
\label{eq:SIR_SD}\vspace*{-1mm}
\end{equation}
Then, we can rewrite the secondary outage probability from \eqref{eq:Psout} as follows:\vspace*{-3mm}

\begin{equation}
P_{\mathrm{s,out}}(\xi_{\mathrm{s}}) = \mathrm{Pr}\left(\min \left(\gamma_{\mathrm{SR}}, \gamma_{\mathrm{SD}}\right) < \xi_{\mathrm{s}} \right),\vspace*{-1mm}
\label{eq:Psout1}
\end{equation}
where $\min \left(\gamma_{\mathrm{SR}}, \gamma_{\mathrm{SD}}\right)$ is the instantaneous end-to-end SIR between ST and SD and $\xi_{\mathrm{s}} = 2^{\frac{2 \mathcal{R}_{\mathrm{s}}}{(1-\alpha)BL}}-1$.  
Using the independence between $\gamma_{\mathrm{SR}}$ and $\gamma_{\mathrm{SD}}$, we can write \eqref{eq:Psout1} as\vspace*{-1mm}
\begin{equation}
P_{\mathrm{s,out}}(\xi_{\mathrm{s}})  = 1 - \big[(1-\underbrace{\mathrm{Pr}(\gamma_{\mathrm{SR}} < \xi_{\mathrm{s}})}_{F_{\mathrm{SR}}(\xi_{\mathrm{s}})})(1-\underbrace{\mathrm{Pr}(\gamma_{\mathrm{SD}} < \xi_{\mathrm{s}})}_{F_{\mathrm{SD}}(\xi_{\mathrm{s}})})\big],\vspace*{-1mm}
\label{eq:Psout_fin1}
\end{equation}
where $F_{\mathrm{SR}}(\xi_{\mathrm{s}})$ and $F_{\mathrm{SD}}(\xi_{\mathrm{s}})$ are the cumulative distribution functions (CDFs) of RVs $\gamma_{\mathrm{SR}}$ and $\gamma_{\mathrm{SD}}$, respectively.\vspace*{-2mm}
\begin{proposition}
The CDF $F_{\mathrm{SR}}(\xi_{\mathrm{s}})$ is\vspace*{-1mm}
\begin{equation}
F_{\mathrm{SR}}(\xi_{\mathrm{s}}) = 1 -\left(1+\frac{P_{\mathrm{PT}}\lambda_{\mathrm{pr}}}{P_{\mathrm{Sm}}\lambda_{\mathrm{sr}}}\xi_{\mathrm{s}}\right)^{-L}.
\label{eq:cdf_SR}\vspace*{-1mm}
\end{equation}
\end{proposition}\vspace*{-1mm}
\begin{proof}
The proof is given in Appendix~\ref{sec:cdf_SR}.
\end{proof}\vspace*{-1mm}

\begin{proposition}
The CDF $F_{\mathrm{SD}}(\xi_{\mathrm{s}})$ is\vspace*{-1mm}
\begin{equation}
F_{\mathrm{SD}}(\xi_{\mathrm{s}}) = \mathcal{I}(1-\mathrm{P}_{\mathcal{H}_1})  +\!\left[ 1 - \!\left(\! 1+\frac{\mathcal{D}}{P_{\mathrm{R}}\lambda_{\mathrm{rd}}}\xi_{\mathrm{s}}\right)^{-L}\right]\mathrm{P}_{\mathcal{H}_1},
\label{eq:cdf_SD}\vspace*{-1mm}
\end{equation}
where \vspace*{-1mm}
\begin{equation}
\mathcal{I} = \frac{2 t^{L}}{\mathcal{B}\mathcal{C}(\mathcal{AD})^{L}}\big[\mathcal{I}_1 - \mathcal{I}_2\big],
\label{eq:I}\vspace*{-1mm}
\end{equation}
with $\mathcal{A} = P_{\mathrm{PT}}\lambda_{\mathrm{pr}}$, $\mathcal{B} = \frac{2\alpha \delta \lambda_{\mathrm{rd}}}{1-\alpha}$, $\mathcal{C} = P_{\mathrm{Sm}}\lambda_{\mathrm{sr}}$, $\mathcal{D} = P_{\mathrm{PT}}\lambda_{\mathrm{pd}}$, and $t = \left(\frac{1}{\mathcal{A}} - \frac{1}{\mathcal{C}}\right)^{-1}$. The term $\mathcal{I}_1$ in \eqref{eq:I} is given as
\begin{equation*}
\mathcal{I}_1 = \frac{\mathcal{B}\mathcal{C}\mathcal{D}^{L}}{2}\left[1 - \Gamma(L+1)\exp\left(\frac{\xi_{\mathrm{s}}\mathcal{D}}{2\mathcal{B}\mathcal{C}}\right)W_{-L,\frac{1}{2}}\left(\frac{\xi_{\mathrm{s}}\mathcal{D}}{\mathcal{B}\mathcal{C}}\right)\right],
\end{equation*}
where $\Gamma(\cdot)$ is the Gamma function~\cite[8.31]{gradshteyn} and $W_{\cdot,\cdot}(\cdot)$ is the Whittaker function~\cite[9.22]{gradshteyn}. The term $\mathcal{I}_2$ in \eqref{eq:I} is given as\vspace*{-4mm}

{{\small
\begin{align*}
\mathcal{I}_2 &= \frac{1}{2}\sum_{j = 0 }^{L-1}\frac{1}{\Gamma(j+1)}\left(\frac{1}{t\mathcal{B}\mathcal{F}}\right)^{j}\xi_{\mathrm{s}}^{\frac{j+2}{2}}\left[\Gamma(j+1)\mathcal{D}^{L}\left(\mathcal{F}\sqrt{\xi_{\mathrm{s}}}\right)^{-j-2} \right. \nonumber \\
&\left. - \frac{\Gamma(L + j+1)}{\xi_{\mathrm{s}}\mathcal{F}^2}\exp\left(\frac{\xi_{\mathrm{s}}\mathcal{D}\mathcal{F}^2}{2}\right)\!\!{\mathcal{D}}^{\frac{2L+j}{2}}W_{-\frac{2L + j}{2}, \frac{j+1}{2}}\!\left(\xi_{\mathrm{s}}\mathcal{D}\mathcal{F}^2\right)\right],
\end{align*}}}\vspace*{-2mm}

\noindent where $\mathcal{F} = \sqrt{\frac{1}{\mathcal{B}}\left(\frac{1}{\mathcal{C}} + \frac{1}{t}\right)}$. The term $\mathrm{P}_{\mathcal{H}_1}$ in \eqref{eq:cdf_SD} is given as\vspace*{-1mm}
\begin{align*}
 \mathrm{P}_{\mathcal{H}_1} &= 1-\frac{1}{(\mathcal{A})^{L}\Gamma({L})}\bigg[\Upsilon\left(L, \frac{(1-\alpha)P_{\mathrm{R}}}{2\alpha \delta{\mathcal{A}}}\right)(\mathcal{A})^{L}\nonumber \\
& -\exp\left(-\frac{(1-\alpha)P_{\mathrm{R}}}{2\alpha \delta \mathcal{C}}\right)t^{L}\Upsilon\left(L, \frac{(1-\alpha)P_{\mathrm{R}}}{2\alpha \delta t}\right)\bigg],
\end{align*}\vspace*{-3mm}

\noindent where $\Upsilon(\cdot,\cdot)$ is the lower incomplete Gamma function~\cite[8.35]{gradshteyn}.
\end{proposition}\vspace*{-2mm}


\begin{proof}
The proof is given in Appendix~\ref{sec:cdf_SD}.
\end{proof}

Finally, simplifying \eqref{eq:Psout_fin1}, we can express the secondary outage probability as\vspace*{-1mm}
\begin{equation}
P_{\mathrm{s,out}}(\xi_{\mathrm{s}})  = F_{\mathrm{SR}}(\xi_{\mathrm{s}})  + F_{\mathrm{SD}}(\xi_{\mathrm{s}})  - F_{\mathrm{SR}}(\xi_{\mathrm{s}}) F_{\mathrm{SD}}(\xi_{\mathrm{s}}).
\label{eq:Psout_fin2}
\end{equation}

\section{Results and Discussions} 
\subsection{System Parameters and Simulation Setup}
We assume the following system parameters: The desired primary rate, $\mathcal{R}_{\mathrm{p}} = \mathrm{0.4}\,\mathrm{bits/s/Hz}$, the desired secondary rate, $\mathcal{R}_{\mathrm{s}} = \mathrm{0.2}\,\mathrm{bits/s/Hz}$, the energy conversion efficiency factor, $\delta = \mathrm{0.5}$, the primary transmit power, $P_{\mathrm{PT}} = 20\,\mathrm{dB}$. We consider a 2-D simulation setup, where ($x_{i}$, $y_{i}$) is the coordinate of $i$th user. The mean channel gain between $i$th and $j$th users is $d_{ij}^{-\rho}$, where $d_{ij}$ is the distance between users $i$ and $j$, and $\rho$ is the path-loss coefficient which is assumed to be $\mathrm{4}$. The ST, SR, and SD are placed at ($\mathrm{0}$, $\mathrm{0}$), ($\mathrm{0.5}$, $\mathrm{0}$), and ($\mathrm{1}$, $\mathrm{0}$), respectively. The PTs are collocated at ($\mathrm{0.5}$, $\mathrm{1}$), while PDs are collocated at ($\mathrm{1}$, $\mathrm{1}$).

\begin{figure}
\centering
\includegraphics[scale=0.38]{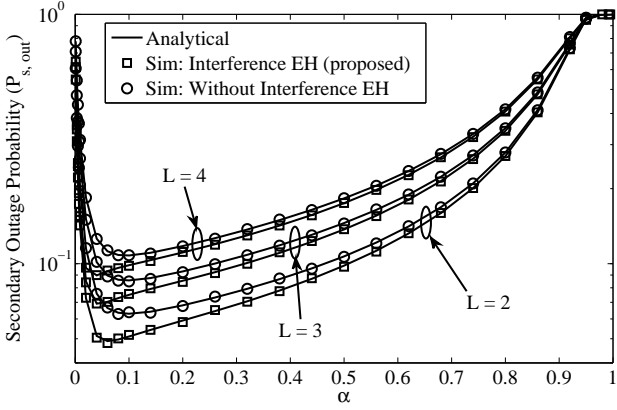}\vspace*{-2mm}
\caption{With interference EH versus without interference EH for different number of primary transceivers ($L$), $\Theta_{\mathrm{p}} = \mathrm{10^{-2}}$, $P_{\mathrm{t}} = \mathrm{20}\,\mathrm{dB}$.}\vspace*{-3mm}
\label{fig:int_no_int}
\end{figure}

\subsection{Effect of the Interference-Assisted Energy Harvesting}
Fig.~\ref{fig:int_no_int} shows SU's outage probability $P_{\mathrm{s,out}}$ against the energy harvesting ratio $\alpha$. We observe that the proposed method of SR harvesting energy from the primary interference in addition to that from the received  information signal, achieves lower $P_{\mathrm{s,out}}$ than the conventional method where SR treats the interference as an unwanted signal in EH phase. This improvement comes from the extraction of an additional energy from the interference, which helps increase the relay's transmit power on SR-SD link, enhancing SIR at SD. For a given number of primary transceivers $L$, as $\alpha$ increases from 0 to 1, $P_{\mathrm{s,out}}$ reduces first, and then increases beyond the optimal value of $\alpha$ that minimizes $P_{\mathrm{s,out}}$. This tradeoff can be attributed to two conflicting effects that are dependent on $\alpha$. The increase in $\alpha$ allows SR to harvest more energy from the information signal and the primary interference, improving SIR of SR-SD link, which in turn, reduces $P_{\mathrm{s,out}}$. On the contrary, the time for data transmission reduces with increasing $\alpha$, which reduces SU's throughput. This pushes SU into the outage more often, increasing its outage probability. Also, we can see that the extra energy gained from the primary interference reduces the optimal value of $\alpha$ as expected.

Similarly, the increase in the number of primary transceivers $L$ furnishes SR with the more harvested energy through the interference, which further reduces the optimal $\alpha$. But, as shown in Fig.~\ref{fig:int_no_int}, the deteriorating effect of the interference$-$decrease in SIR at both SR and SD$-$is more dominant, which increases $P_{\mathrm{s,out}}$. An another negative consequence of the increase in $L$ is the stricter primary outage constraint. Since SU should satisfy the outage constraint of each PU, the increase in the number of PUs makes the constraint more difficult to satisfy, reducing the maximum allowed transmit powers for both ST and SD.\vspace*{-1mm}



\begin{figure}
\centering
\includegraphics[scale=0.38]{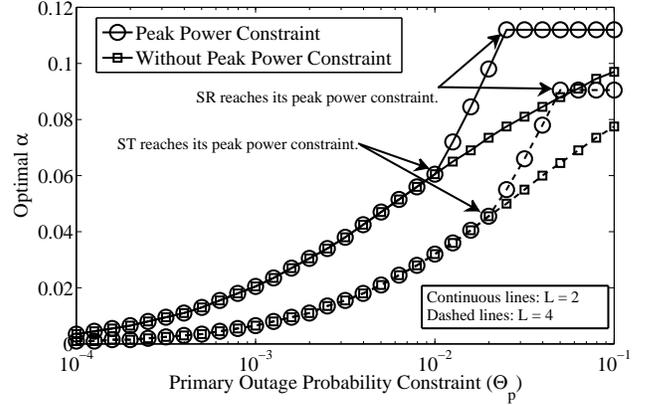}\vspace*{-2mm}
\caption{Optimal $\alpha$ versus the primary outage constraint for different number of primary transceivers ($L$), $P_{\mathrm{t}} = \mathrm{20}\,\mathrm{dB}$.}\vspace*{-3mm}
\label{fig:int_no_int1}
\end{figure}

\subsection{Effect of the Primary Outage Constraint}
\begin{figure}
\centering
\includegraphics[scale=0.38]{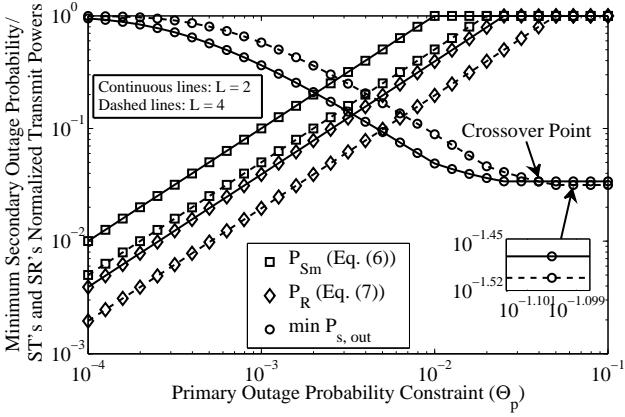}\vspace*{-2mm}
\caption{Minimum $P_{\mathrm{s,out}}$ versus the primary outage constraint with peak power constraint for different number of primary transceivers ($L$), $P_{\mathrm{t}} = \mathrm{20}\,\mathrm{dB}$.}\vspace*{-3mm}
\label{fig:int_no_int2}
\end{figure}

Figs.~\ref{fig:int_no_int1}  and \ref{fig:int_no_int2} show the effect of the primary outage constraint ($\Theta_{\mathrm{p}}$) on the optimal $\alpha$ and its corresponding minimum $P_{\mathrm{s,out}}$, respectively, for different $L$ and peak power constraint of $P_{\mathrm{t}}$. From \eqref{eq:cdf_SR}, \eqref{eq:cdf_SD}, and \eqref{eq:Psout_fin2}, we can see that, deriving the analytical expression for the optimal $\alpha$ is difficult due to the involvement of Whittaker function and incomplete Gamma function in an intricate manner; however, the optimal $\alpha$ can be easily obtained numerically. We note from Fig.~\ref{fig:int_no_int1} that, relaxing the primary outage constraint $\Theta_{\mathrm{p}}$ increases the optimal $\alpha$. This is because, relaxing $\Theta_{\mathrm{p}}$ allows ST and SR to transmit with higher powers. Thus, $\alpha$ increases to cater relay's higher transmit power. Also, higher transmit powers of ST and SR increases SIR on both ST-SR and SR-SD links, which provides an extra margin to increase $\alpha$ improving SU's outage performance. 

The peak power constraint becomes active due to the increased maximum allowed powers for ST ($P_{\mathrm{ST}}$, \eqref{eq:PST}) and SR ($P_{\mathrm{SR}}$, \eqref{eq:Ppout7}) with the relaxation of $\Theta_{\mathrm{p}}$ beyond a threshold.
This is seen in Fig.~\ref{fig:int_no_int1}, where ST reaches its peak power constraint first{\footnote{In simulation setup, ST is located farther from the primary destinations than SR. This allows ST to transmit with higher power than that of SR for the same $\Theta_{\mathrm{p}}$, causing ST to reach the peak power constraint before SR. For the purpose of exposition, the effect of distances among nodes is not addressed in this paper.}} which forces ST to transmit with peak power $P_{\mathrm{t}}$ even though the further relaxation of $\Theta_{\mathrm{p}}$ allows it to transmit with higher power. After this point, to serve the increasing SR's transmit power for a fixed ST's power $P_{\mathrm{t}}$, the optimal $\alpha$ increases at a faster rate than that without the peak power constraint till the peak power constraint of SR is reached. Once SR's peak power constraint is reached, SR is also forced to transmit with the fixed power $P_{\mathrm{t}}$ for any further increase in $\Theta_{\mathrm{p}}$, and the optimal $\alpha$ remains the same thereafter. 

As aforementioned, the increase in $L$ reduces the maximum allowed power for both ST and SR, which delays the arrival of the peak power constraint as shown in Fig.~\ref{fig:int_no_int1}. This has an interesting consequence on the minimum $P_{\mathrm{s,out}}$ as shown in Fig.~\ref{fig:int_no_int2}. At the stringent $\Theta_{\mathrm{p}}$, for lower $L$ ($L = 2$), the minimum $P_{\mathrm{s,out}}$ is lower than that for higher $L$ ($L = 4$). However, there exists a crossover point, after which the trend reverses; because, for $L = 2$, the peak power constraint is reached for both ST and SR earlier, forcing them to transmit with fixed power $P_{\mathrm{t}}$ even with the further relaxation of $\Theta_{\mathrm{p}}$. Meanwhile, for $L = 4$, more energy is harvested from the interference than for $L = 2$, and ST and SR may keep increasing their transmit powers even at $\Theta_{\mathrm{p}}$ for which the peak power constraint for $L = 2$ is reached, allowing the former to achieve a better minimum $P_{\mathrm{s,out}}$ at higher $\Theta_{\mathrm{p}}$. Note that we do not observe such behavior in Fig.~\ref{fig:int_no_int}, as for $\Theta_{\mathrm{p}} = 10^{-2}$ as assumed in it, the peak power constraint is not reached for $L = 2, 3,$ and $4$. Combining both the primary outage constraint and the peak power constraint, Fig.~\ref{fig:int_no_int2} has plotted the maximum allowed transmit powers for ST and SR normalized by their peak power constraint power $P_\mathrm{t}$.\vspace*{-1mm}

\section{Conclusions}
In this paper, we have considered the spectrum sharing of the secondary system with multiple primary transceivers, where the secondary users communicate via an energy harvesting decode-and-forward relay under the primary outage constraint. The secondary relay harvests energy from the received information signal as well as from the primary interference, which is used to forward the data to the secondary destination. We have adopted the time switching protocol which allows the relay to switch between the energy harvesting and the information processing.

For the proposed scenario, we have derived a closed-form expression for the secondary outage probability. We have shown that, harvesting energy from the primary interference achieves a better secondary outage performance and reduces the optimal value of the energy harvesting ratio $\alpha$. Though the increase in the number of primary transceivers reduces the optimal value of $\alpha$ further, it increases the minimum secondary outage probability when the peak power constraint is inactive. Interestingly, the trend reverses for the minimum secondary outage probability, once the peak power constraint becomes active with the relaxation of the primary outage constraint.

%

\appendices
\section{Proof of \eqref{eq:PST}}
\label{sec:PST}
Let $\mathcal{K}$ be $\mathrm{Pr}\left( \frac{P_{\mathrm{PT}}|h_{\mathrm{pp}_i}|^{2}}{P_{\mathrm{ST}}|h_{\mathrm{sp}_i}|^{2}} \leq \zeta_{\mathrm{p}}\right) $. Then, we can write\vspace*{-1mm} 
\begin{align}
\mathcal{K} = \int_{0}^{\infty} \mathrm{Pr}\left( \frac{P_{\mathrm{PT}}|h_{\mathrm{pp}_i}|^{2}}{P_{\mathrm{ST}}y} \leq \zeta_{\mathrm{p}}\right)f_{|h_{\mathrm{sp}_i}|^{2}}(y)\mathrm{d}y,
\label{eq:Ppout3}
\end{align}\vspace*{-4mm}

\noindent where $f_{|h_{\mathrm{sp}_i}|^{2}}(y)$ is the probability density function of $|h_{\mathrm{sp}_i}|^{2}$, and is given by $f_{|h_{\mathrm{sp}_i}|^{2}}(y) = \frac{1}{\lambda_{\mathrm{sp}}}\exp\left(-\frac{y}{\lambda_{\mathrm{sp}}}\right)$. Solving \eqref{eq:Ppout3} and then substituting the value of $\mathcal{K}$ in \eqref{eq:Ppout2}, we obtain\vspace*{-1mm} 
\begin{align}
\mathrm{P}_{\mathrm{p, out, ST}}  = 1 - \left(\frac{P_{\mathrm{PT}}\lambda_{\mathrm{pp}}}{P_{\mathrm{ST}}\lambda_{\mathrm{sp}}\zeta_{\mathrm{p}} + P_\mathrm{PT}\lambda_{\mathrm{pp}}}\right)^{L}.
\label{eq:Ppout5}
\end{align}\vspace*{-3mm} 

\noindent Solving \eqref{eq:Ppout5} for $P_{\mathrm{ST}}$, we obtain the required expression in \eqref{eq:PST}.

\section{Proof of \eqref{eq:cdf_SR}}
\label{sec:cdf_SR}\vspace*{-1mm}
Let us write $\gamma_{\mathrm{SR}}$ from \eqref{eq:SIR_SR} as\vspace*{-2mm} 
\begin{equation}
\gamma_{\mathrm{SR}} = \frac{X}{Y},
\label{eq:A1}\vspace*{-1mm} 
\end{equation}
where $X = P_{\mathrm{Sm}} |h_{\mathrm{sr}}|^{2}$ is the exponentially distributed RV with mean $\lambda_{\mathrm{x}} = P_{\mathrm{Sm}}\lambda_{\mathrm{sr}}$ with the probability density function (PDF) given by $f_{X}(x) = \frac{1}{\lambda_{\mathrm{x}}}\exp\left(-\frac{x}{\lambda_{\mathrm{x}}}\right)$  and $Y = \displaystyle \sum_{i = 1}^{L}P_{\mathrm{PT}}|h_{\mathrm{p}_i \mathrm{r}}|^2$ is the Gamma distributed RV with a shape parameter $L$ and a scale parameter $\lambda_{\mathrm{y}}$, and its PDF is given by $f_{Y}(y) = \frac{1}{\lambda_{\mathrm{y}}^{L}\Gamma(L)}y^{L-1}\exp\left(-\frac{y}{\lambda_{\mathrm{y}}}\right)$, where $\lambda_{\mathrm{y}} = P_{\mathrm{PT}}\lambda_{\mathrm{pr}}$. Thus, we can write CDF of $\gamma_{\mathrm{SR}}$ as\vspace*{-4mm}

{{\small\begin{align}
F_{\mathrm{SR}}(\xi_{\mathrm{s}}) &= \mathrm{Pr}\left(\frac{X}{Y} < \xi_{\mathrm{s}}\right) \nonumber \\
&\hspace*{-10mm} =\frac{1}{\lambda_{\mathrm{x}}\lambda_{\mathrm{y}}^{L}\Gamma(L)}\displaystyle\! \int_{y = 0}^{\infty} \!\displaystyle\int_{x = 0}^{\xi_{\mathrm{s}}y}\!\! \exp\left(-\frac{x}{\lambda_{\mathrm{x}}}\right)\! y^{L-1}\!\! \exp\!\left(\!-\frac{y}{\lambda_{\mathrm{y}}}\right) \mathrm{d}x\,\mathrm{d}y,
\label{eq:int}
\end{align}}}\vspace*{-3mm}

\noindent where we have used the independence between $|h_{\mathrm{sr}}|^{2}$ and $|h_{\mathrm{p}_i \mathrm{r}}|^2$. Solving \eqref{eq:int}, we obtain the required expression in \eqref{eq:cdf_SR}.\vspace*{-1mm}

\section{Proof of \eqref{eq:cdf_SD}}
\label{sec:cdf_SD}\vspace*{-1mm}

We write $\gamma_{\mathrm{SD}}$ from \eqref{eq:harv_power}, \eqref{eq:SR_fin}, and \eqref{eq:SIR_SD} as\vspace*{-1mm} 
\begin{align}
\gamma_{\mathrm{SD}} &= \frac{\min\left({P_{\mathrm{SR,H}}}, P_{\mathrm{R}}\right) |h_{\mathrm{rd}}|^{2}}{\sum_{i = 1}^{L}P_{\mathrm{PT}}|h_{\mathrm{p}_i \mathrm{d}}|^2},
\label{eq:1}
\end{align}\vspace*{-3mm} 

\noindent where ${P_{\mathrm{SR,H}}}$ is given by \eqref{eq:harv_power}. Let us denote $G_1 = \sum_{i=1}^{L}\!P_{\mathrm{PT}}|h_{\mathrm{p}_i\mathrm{r}}|^2$, $G_2= P_{\mathrm{Sm}}|h_{\mathrm{sr}}|^2$, $Z_1 =\sum_{i=1}^{L} P_{\mathrm{PT}}|h_{\mathrm{p}_i\mathrm{d}}|^2$, and $Z_2 =2\delta \alpha |h_{\mathrm{rd}}|^2/(1-\alpha)$. Then, $G_1$ and $Z_1$ are Gamma distributed RVs with the same shape parameter $L$ and a scale parameters $\mathcal{A} = P_{\mathrm{PT}}\lambda_{\mathrm{pr}}$ and $\mathcal{D} = P_{\mathrm{PT}}\lambda_{\mathrm{pd}}$, respectively; while $G_2$ and $Z_2$ are exponentially distributed RVs with means $\mathcal{C} = P_{\mathrm{Sm}}\lambda_{\mathrm{sr}}$ and $\mathcal{B} = 2\delta \alpha\lambda_{\mathrm{rd}}/(1-\alpha)$, respectively. Then, we can write the CDF of $\gamma_{\mathrm{SD}}$ in \eqref{eq:1} as\vspace*{-3mm}

\begin{align}
F_{\mathrm{SD}}(\xi_{\mathrm{s}}) & = \underbrace{\mathrm{Pr}\left(\frac{ (1-\alpha)P_{\mathrm{R}} Z_2}{2\delta \alpha Z_1} \leq \xi_{\mathrm{s}}\right)}_{\mathcal{J}}\underbrace{\mathrm{Pr}\left( P_{\mathrm{SR,H}} \geq P_{\mathrm{R}}\right)}_{\mathrm{P}_{\mathcal{H}_1}} \nonumber\\
&+ \underbrace{\mathrm{Pr}\left(\frac{Z_2\left(G_1+G_2\right)}{Z_1} \leq \xi_{\mathrm{s}}\right)}_{\mathcal{I}}\underbrace{\mathrm{Pr}\left( P_{\mathrm{SR,H}} < P_{\mathrm{R}}\right)}_{1-\mathrm{P}_{\mathcal{H}_1}}.
\label{eq:Bmain}
\end{align}
Following the steps to derive \eqref{eq:cdf_SR}, we can write $\mathcal{J}$ as
\begin{equation}
\mathcal{J} =  1 -\left(1+\frac{\mathcal{D}}{P_{\mathrm{R}}\lambda_{\mathrm{rd}}}\xi_{\mathrm{s}}\right)^{-L}.
\label{eq:B1}\vspace*{-1mm}
\end{equation}
We can write $\mathrm{P}_{\mathcal{H}_1}$ as $
\mathrm{P}_{\mathcal{H}_1} = \mathrm{Pr}\left(G_1 + G_2 \geq P_{\mathrm{R}}^{*}\right)$, where $P_{\mathrm{R}}^{*} = \frac{(1-\alpha)P_{\mathrm{R}}}{2\alpha \delta}$. Using the independence between $|h_{\mathrm{p}_i\mathrm{r}}|^2$ and $|h_{\mathrm{sr}}|^2$, $\mathrm{P}_{\mathcal{H}_1}$ can be written as\vspace*{-2mm}
\begin{align}
\mathrm{P}_{\mathcal{H}_1} \!=\! 1\!-\!\frac{1}{\mathcal{C}\mathcal{A}^{L}\Gamma(L)}\!\! \int_{0}^{P_{\mathrm{R}}^{*}}\!\!\!\!\int_{g_2=0}^{P_{\mathrm{R}}^{*}-g_1}\!\!\!\!\!\!\exp\!\left(\!\!-\frac{g_2}{\mathcal{C}}\!\right)\!g_1^{L-1}\!\exp\!\left(\!\!-\frac{g_1}{\mathcal{A}}\!\right)\!\mathrm{d}g_2\mathrm{d}g_1.
\label{eq:B2}
\end{align}\vspace*{-4mm}

\noindent Solving \eqref{eq:B2}, we get\vspace*{-1mm}
\begin{align}
 \mathrm{P}_{\mathcal{H}_1} &= 1-\frac{1}{(\mathcal{A})^{L}\Gamma({L})}\bigg[\Upsilon\left(L, \frac{(1-\alpha)P_{\mathrm{R}}}{2\alpha \delta{\mathcal{A}}}\right)(\mathcal{A})^{L}\nonumber \\
& -\exp\left(-\frac{(1-\alpha)P_{\mathrm{R}}}{2\alpha \delta \mathcal{C}}\right)t^{L}\Upsilon\left(L, \frac{(1-\alpha)P_{\mathrm{R}}}{2\alpha \delta t}\right)\bigg].
\label{eq:B3}
\end{align}\vspace*{-3mm}

\noindent Denote $Z = G_1 + G_2$. Then, we can write PDF of $Z$ as follows:\vspace*{-1mm}
\begin{align}
f_{Z}(z) &=  \frac{1}{\mathcal{C}\mathcal{A}^{L}\Gamma(L)}\int_{0}^{z}\!\!\exp\!\left(\!\!-\frac{z-g_1}{\mathcal{C}}\!\right)\!g_1^{L-1}\!\exp\!\left(\!\!-\frac{g_1}{\mathcal{A}}\!\right)\mathrm{d}g_1\nonumber \\
 &= \frac{t^L}{\Gamma(L)\mathcal{A}^L \mathcal{C}}\exp\left(-\frac{z}{\mathcal{C}}\right)\Upsilon\left(L,\frac{z}{t}\right),
\label{eq:B5}\vspace*{-1mm}
\end{align}
where $t = \left(\frac{1}{\mathcal{A}} - \frac{1}{\mathcal{C}}\right)^{-1}$. We denote $Q = Z_2(G_1 + G_2) = Z_2 Z$. Then, we can write PDF of $Q$ as follows:\vspace*{-1mm}
\begin{align}
f_{Q}({q}) \!=\! \frac{t^L\mathcal{A}^{-L}}{\Gamma(L)\mathcal{B} \mathcal{C}}\int_{0}^{\infty}\!\!\!\exp \!\left(\! -\frac{z_2}{\mathcal{B}}\right)\! \exp \! \left(\!-\frac{q}{z_2 \mathcal{C}}\! \right)\!\Upsilon \! \left(\! L,\frac{q}{z_2t}\! \right)\!\frac{1}{z_2}\mathrm{d}z_2.
\label{eq:B6}
\end{align}\vspace*{-4mm}

\noindent Since $L$ takes positive integer values, we use the series expansion of lower incomplete Gamma function $\Upsilon(a, b)$ for positive integer values of $a$ as $(a-1)!\left(1-\exp(-b)\sum_{k=0}^{a-1}\frac{b^k}{k!}\right)$. Also, using \cite[3.471.12]{gradshteyn}, we can express \eqref{eq:B6} as
\begin{align}
f_{Q}({q}) \!=\! \frac{2t^L}{\mathcal{B}\mathcal{C}\mathcal{A}^{L}}\!\!\left[\!K_{0}\!\left(\!2\sqrt{\frac{q}{\mathcal{B}\mathcal{C}}}\!\right) \!-\!\sum_{j=0}^{L-1}\!\frac{(\mathcal{B\theta)}^{-\frac{j}{2}}\left(q\right)^{\frac{j}{2}}}{{t}^{j}\Gamma(j+1)}K_{j} \! \left(\!\! 2\sqrt{\frac{q\theta}{\mathcal{B}}}\right)\!\right]\!,
\end{align}\vspace*{-3mm}

\noindent where $K_{\nu}(\cdot)$ is the modified Bessel function of second kind~\cite[8.43]{gradshteyn} and $\theta = \frac{1}{\mathcal{C}} + \frac{1}{t}$. Then, we can write $\mathcal{I}$ in \eqref{eq:Bmain} as\vspace*{-1mm}
\begin{align}
\mathcal{I} &= \frac{2t^L}{\mathcal{B}\mathcal{C}(\mathcal{AD})^L\Gamma(L)}\int_{z_1 = 0}^{\infty} \int_{q = 0}^{z_1{\xi_{\mathrm{s}}}}\left[\!K_{0}\!\left(\!2\sqrt{\frac{q}{\mathcal{B}\mathcal{C}}}\right)\right. \nonumber \\
& \left.-\!\sum_{j=0}^{L-1}\!\frac{(\mathcal{B\theta)}^{-\frac{j}{2}}\left(q\right)^{\frac{j}{2}}}{{t}^{j}\Gamma(j+1)}K_{j} \! \left(\!\! 2\sqrt{\frac{q\theta}{\mathcal{B}}}\right)\!\right]\!z_1^{L-1}\exp\!\left(\!-\frac{z_1}{\mathcal{D}}\! \right)\mathrm{d}q\,\mathrm{d}z_1.
\end{align}\vspace*{-4mm}

\noindent Using \cite[6.561.8]{gradshteyn}, we obtain the required $\mathcal{I}$ in \eqref{eq:cdf_SD} in closed-form as \eqref{eq:I}. Substituting $\mathcal{I}$ along with $\mathcal{J}$ from \eqref{eq:B1} and $\mathrm{P}_{\mathcal{H}_1}$ from \eqref{eq:B2} in \eqref{eq:Bmain}, we get the required closed-form expression of CDF of $\gamma_{\mathrm{SD}}$ as in \eqref{eq:cdf_SD}.\vspace*{-2mm}

\bibliographystyle{ieeetr}
\bibliography{paper}

\begin{thebibliography}{10}

\bibitem{sultan}
A.~Sultan, ``Sensing and transmit energy optimization for an energy harvesting
  cognitive radio,'' {\em IEEE Wireless Commun. Lett.}, vol.~1, no.~5,
  pp.~500--503, Oct. 2012.

\bibitem{lee1}
S.~Lee, R.~Zhang, and K.~Huang, ``Opportunistic wireless energy harvesting in
  cognitive radio networks,'' {\em IEEE Trans. Wireless Commun.}, vol.~12,
  no.~9, pp.~4788--4799, Sept. 2013.

\bibitem{jeya}
J.~Jeya~Pradha, S.~S. Kalamkar, and A.~Banerjee, ``Energy harvesting cognitive
  radio with channel-aware sensing strategy,'' {\em IEEE Commun. Lett.},
  vol.~18, no.~7, pp.~1171--1174, July 2014.

\bibitem{usman}
M.~Usman and I.~Koo, ``Access strategy for hybrid underlay-overlay cognitive
  radios with energy harvesting,'' {\em IEEE Sensors J.}, vol.~14, no.~9,
  pp.~3164--3173, Sept. 2014.

\bibitem{shaf}
A.~E. Shafie, M.~Ashour, T.~Khattab, and A.~Mohamed, ``On spectrum sharing
  between energy harvesting cognitive radio users and primary users,'' in {\em
  Proc. 2015 ICNC}, pp.~214--220.

\bibitem{lu}
X.~Lu, P.~Wang, D.~Niyato, D.~I. Kim, and Z.~Han, ``Wireless networks with {RF}
  energy harvesting: {A} contemporary survey,'' {\em IEEE Commun. Surveys
  Tuts.}, vol.~17, no.~2, Second Quarter 2015.

\bibitem{varshney}
L.~R. Varshney, ``Transporting information and energy simultaneously,'' in {\em
  Proc. 2008 IEEE ISIT}, pp.~1612--1616.

\bibitem{grover}
P.~Grover and A.~Sahai, ``{Shannon} meets {Tesla}: {W}ireless information and
  power transfer,'' in {\em Proc. 2010 IEEE ISIT}, pp.~2363--2367.

\bibitem{rui2}
R.~Zhang and C.~K. Ho, ``{MIMO} broadcasting for simultaneous wireless
  information and power transfer,'' {\em IEEE Trans. Wireless Commun.},
  vol.~12, no.~5, pp.~1989--2001, May 2013.

\bibitem{rui3}
L.~Liu, R.~Zhang, and K.-C. Chua, ``Wireless information transfer with
  opportunistic energy harvesting,'' {\em IEEE Trans. Wireless Commun.},
  vol.~12, no.~1, pp.~288--300, Jan. 2013.

\bibitem{nasir}
A.~A. Nasir, X.~Zhou, S.~Durrani, and R.~A. Kennedy, ``Relaying protocols for
  wireless energy harvesting and information processing,'' {\em IEEE Trans.
  Wireless Commun.}, vol.~12, pp.~3622--3636, July 2013.

\bibitem{aissa}
Y.~Gu and S.~Aissa, ``Interference aided energy harvesting in
  decode-and-forward relaying systems,'' in {\em Proc. 2014 IEEE ICC},
  pp.~5378--5382.

\bibitem{krik}
I.~Krikidis, S.~Timotheou, and S.~Sasaki, ``{RF} energy transfer for
  cooperative networks: {D}ata relaying or energy harvesting?,'' {\em IEEE
  Commun. Lett.}, vol.~16, no.~11, pp.~1772--1775, Nov. 2012.

\bibitem{yener1}
K.~Tutuncuoglu and A.~Yener, ``Cooperative energy harvesting communications
  with relaying and energy sharing,'' in {\em Proc. 2013 IEEE ITW}, pp.~1--5.

\bibitem{ishibashi1}
K.~Ishibashi, ``Dynamic harvest-and-forward: New cooperative diversity with
  {RF} energy harvesting,'' in {\em Proc. 2014 WCSP}, pp.~1--5.

\bibitem{poor}
Z.~Ding, S.~M. Perlaza, I.~Esnaola, and H.~V. Poor, ``Power allocation
  strategies in energy harvesting wireless cooperative networks,'' {\em IEEE
  Trans. Wireless Commun.}, vol.~13, no.~2, pp.~846--860, Feb. 2014.

\bibitem{gan}
G.~Zheng, Z.~Ho, E.~A. Jorswieck, and B.~Ottersten, ``Information and energy
  cooperation in cognitive radio networks,'' {\em IEEE Trans. Signal Process.},
  vol.~62, no.~9, pp.~2290--2303, May 2014.

\bibitem{micha}
D.~S. Michalopoulos, H.~A. Suraweera, and R.~Schober, ``The impact of relay
  selection on the tradeoff between information transmission and wireless
  energy transfer,'' in {\em Proc. 2014 IEEE GLOBECOM}, pp.~4191--4196.

\bibitem{nasir2}
A.~A. Nasir, X.~Zhou, S.~Durrani, and R.~A. Kennedy, ``Wireless-powered relays
  in cooperative communications: {T}ime-switching relaying protocols and
  throughput analysis,'' {\em IEEE Trans. Commun.}, vol.~63, no.~5,
  pp.~1607--1622, May 2015.

\bibitem{chen1}
H.~Chen, Y.~Li, J.~L. Rebelatto, B.~F. Uchoa-Filho, and B.~Vucetic,
  ``Harvest-then-cooperate: Wireless-powered cooperative communications,'' {\em
  IEEE Trans. Signal Process.}, vol.~63, no.~7, pp.~1700--1711, Apr. 2015.

\bibitem{van}
V.-D. Nguyen, S.~Dinh-Van, and O.-S. Shin, ``Opportunistic relaying with
  wireless energy harvesting in a cognitive radio system,'' in {\em Proc. 2015
  IEEE WCNC}, pp.~87--92.

\bibitem{mousa}
S.~Mousavifar, Y.~Liu, C.~Leung, M.~Elkashlan, and T.~Duong, ``Wireless energy
  harvesting and spectrum sharing in cognitive radio,'' in {\em Proc. 2014 IEEE
  VTC-Fall}, pp.~1--5.

\bibitem{sanket}
S.~S. Kalamkar and A.~Banerjee, ``Outage analysis of spectrum sharing energy
  harvesting cognitive relays in {N}akagami-$m$ channels,'' in {\em Proc. 2015
  IEEE GLOBECOM}.

\bibitem{he}
H.~Chen, Y.~Li, Y.~Jiang, Y.~Ma, and B.~Vucetic, ``Distributed power splitting
  for {SWIPT} in relay interference channels using game theory,'' {\em IEEE
  Trans. Wireless Commun.}, vol.~14, no.~1, pp.~410--420, Jan. 2015.

\bibitem{zou}
Y.~Zou, J.~Zhu, B.~Zheng, and Y.-D. Yao, ``An adaptive cooperation diversity
  scheme with best-relay selection in cognitive radio networks,'' {\em IEEE
  Trans. Signal Process.}, vol.~58, no.~10, pp.~5438--5445, Oct. 2010.

\bibitem{peter:2013}
P.~J. Smith, P.~A. Dmochowski, H.~A. Suraweera, and M.~Shafi, ``The effects of
  limited channel knowledge on cognitive radio system capacity,'' {\em IEEE
  Trans. Veh. Technol.}, vol.~62, no.~2, pp.~927--933, Feb. 2013.

\bibitem{duong2}
T.~Duong, P.~L. Yeoh, V.~N.~Q. Bao, M.~Elkashlan, and N.~Yang, ``Cognitive
  relay networks with multiple primary transceivers under spectrum-sharing,''
  {\em IEEE Signal Process. Lett.}, vol.~19, no.~11, pp.~741--744, Nov. 2012.

\bibitem{gradshteyn}
I.~S. Gradshteyn and I.~M. Ryzhik, {\em Table of Integrals, Series, and
  Products}.
\newblock Academic Press, {8th}~ed., 2015.

\end{thebibliography}

\end{document}